\documentclass[12pt]{iopart}

\usepackage{graphicx}

\begin{document}
\title[PIC simulation of an expanding plasma cloud]{PIC simulation of 
a thermal anisotropy-driven Weibel instability in a circular rarefaction 
wave}

\author{M. E. Dieckmann\textsuperscript{1}, G. Sarri\textsuperscript{1},  
G. C. Murphy\textsuperscript{2}, A. Bret\textsuperscript{3}, 
L. Romagnani\textsuperscript{4}, I. Kourakis\textsuperscript{1}, 
M. Borghesi\textsuperscript{1}, A. Ynnerman\textsuperscript{5}, 
L. O'C. Drury\textsuperscript{2}}

\address{1 Centre for Plasma Physics, The Queen's University of Belfast, 
Belfast, BT7 1NN, Northern Ireland}
\address{2 Dublin Institute for Advanced Studies, 31 Fitzwilliam Place, 
Dublin 2, Ireland}
\address{3 ETSI Industriales, Universidad de Castilla-La Mancha, 13071 Ciudad Real, 
Spain and Instituto de Investigaciones Energticas y Aplicaciones Industriales, 
Campus Universitario de Ciudad Real, 13071 Ciudad Real, Spain}
\address{4 LULI, Ecole Polytechnique, CNRS, CEA, UPMC; 91128 Palaiseau, France}
\address{5 MIT, Dept of Science and Technology (ITN), Link\"oping University, 
SE-60174 Norrkoping, Sweden}

\date{\today}

\begin{abstract}
The expansion of an initially unmagnetized planar rarefaction wave has recently been 
shown to trigger a thermal anisotropy-driven Weibel instability (TAWI), which can 
generate magnetic fields from noise levels. It is examined here if the TAWI can also 
grow in a curved rarefaction wave. The expansion of an initially unmagnetized circular 
plasma cloud, which consists of protons and hot electrons, into a vacuum is modelled 
for this purpose with a two-dimensional particle-in-cell (PIC) simulation. It is shown 
that the momentum transfer from the electrons to the radially accelerating protons can 
indeed trigger a TAWI. Radial current channels form and the aperiodic growth of a 
magnetowave is observed, which has a magnetic field that is oriented orthogonal to the 
simulation plane. The induced electric field implies that the electron density gradient 
is no longer parallel to the electric field. Evidence is presented here for that this 
electric field modification triggers a second magnetic instability, which results in a 
rotational low-frequency magnetowave. The relevance of the TAWI is discussed for the 
growth of small-scale magnetic fields in astrophysical environments, which are needed 
to explain the electromagnetic emissions by astrophysical jets. It is outlined how this 
instability could be examined experimentally.
\end{abstract}

\pacs{52.80.Qj, 52.65.Rr, 52.35.Tc, 52.35.Qz}
\maketitle

\section{Introduction}

Energetic electromagnetic radiation is emitted by a wide range of astrophysical objects, 
such as supernova remnants (SNRs) \cite{SNRA}, microquasars \cite{Fender} and gamma ray 
bursts (GRBs) \cite{Piran}. This radiation is at least partially attributed to the 
synchrotron- or synchrotron jitter emissions \cite{MedvedevJit} of ultra-relativistic 
electrons, which gyrate in strong magnetic fields. Such extreme conditions do not exist in 
the interstellar medium and probably not in astrophysical jets or in SNR blast shells, if
their plasma would be in a steady state. Efficient electron acceleration and magnetic field 
amplification mechanisms must be at work and they are driven primarily by collisionless 
plasma shocks \cite{Bykov}.

Through their interaction with the electromagnetic fields in the shock transition layer,
the electrons may gain enough energy to be injected into the diffusive shock acceleration 
process \cite{Blandford}. The latter is probably capable of accelerating charged particles 
to the highest observed cosmic ray energies and it should supply a population of highly
relativistic electrons. The apparent magnetization of SNR shocks \cite{Volk} is attributed 
to cosmic ray-driven instabilities, which can operate on large scales \cite{Bell}. 
Filamentation instabilities of counterstreaming beams of charged particles are thought to 
be the origin of the micro-scale magnetic fields, in particular of those in GRB jets 
\cite{Medvedev,Brainerd}, although the lifetimes of these fields might be too short 
\cite{Gruzinov,Waxman}.  

Filamentation or beam-Weibel instabilities develop in an initially spatially uniform 
system of two counter-streaming particle beams. Typically one examines this instability 
in the reference frame, in which the net charge and current vanish. The beams consist 
of electrons and ions and they may contain a sizeable fraction of positrons in GRB jets. 
The filamentation instability transforms the spatially uniform beams into a system of 
spatially separated current flux tubes, which are enwrapped and separated by 
electromagnetic fields [13-17].
Observations of 
the filamentation instability in laboratory settings and the therefrom resulting 
electromagnetic radiation exist \cite{Kapetanakos} and they are likely to occur close 
to shocks \cite{Bulanov} in astrophysical settings.

The Weibel instability \cite{Weibel} is another magnetic instability. It is driven by a 
thermal anisotropy of a single electron species. Weibel considered in his original work a spatially uniform bi-Maxwellian 
electron velocity distribution. The electrons were cooler along one direction than along
the other two. The Weibel instability gives rise to the aperiodic growth of magnetowaves with 
a wavevector that is parallel to the cool direction \cite{Kalman} and with wavelengths that 
are comparable to an electron skin depth. The Weibel instability has been widely examined in 
the past both analytically and numerically [21-30]. Its importance for the generation of 
magnetic fields in jet outflows and on a cosmological scale has been pointed out \cite{Schlickeiser2}. It has, however, previously been unclear how a strong thermal anisotropy 
could form and survive over spatial scales that exceed the wavelength of the modes, which
are destabilized by the Weibel instability in its original form \cite{Weibel}.  

Such a process has been identified recently \cite{Thaury}. A plasma density gradient 
together with the high electron mobility gives rise to an ambipolar electrostatic field, 
which confines the electrons and accelerates the ions. The electric field vector is 
antiparallel to the density gradient. The ion acceleration launches a rarefaction wave [33-38].
The electrons, which move 
from the dense plasma to the front of the rarefaction wave, are slowed down and reflected 
inelastically by the electric field of the expanding wave, through which they provide the 
energy for the ion acceleration. The modulus of the electron momentum along the density 
gradient is reduced by these processes, while both orthogonal momentum components are 
unchanged. The thermal anisotropy can give rise to the Weibel instability. In what follows 
we refer to it as the thermal anisotropy-driven Weibel instability (TAWI). Particle-in-cell 
(PIC) simulations \cite{Thaury} have shown that the TAWI can magnetize a planar plasma 
cloud that expands into a vacuum. A shock develops, if the rarefaction wave expands into 
a dilute ambient medium. The TAWI is suppressed if a shock is limiting the extent of the 
rarefaction wave to a length that is shorter than the wavelength of its modes 
\cite{SarriPRL,SarriNJP}. 

This process, which requires ion density gradients and hot electrons, can 
be important for energetic astrophysical plasma. The internal shock model for the prompt 
GRB emissions assumes that the jet consists of plasma clouds with varying densities and 
mildly relativistic relative speeds \cite{Piran}. We can also assume that the ion density 
close to SNR shocks will not be spatially uniform.  The bulk electrons in the GRB jets 
are thought to be relativistically hot and the bulk electrons close to SNR shocks have 
keV temperatures. The magnetic energy generated by the TAWI can reach a sizeable fraction of the electron 
thermal energy, which could result in strong magnetic fields. 

The TAWI has previously been examined numerically in a two-dimensional planar geometry. 
A plasma slab is centred in these studies in the simulation box and confined by two 
parallel boundaries, across which the density jumps from the constant high density within 
the slab to a low density or a vacuum outside of it. The plasma density is constant along 
the second simulation direction and the boundary conditions are periodic. The rarefaction 
waves launched at both boundaries move into opposite directions, causing an expansion of 
the slab plasma. The density gradients of both rarefaction waves are antiparallel. An 
electron can cross the slab many times and its momentum component along the density 
gradients remains decoupled from both orthogonal ones. Multiple electron reflections, 
which reduce only the electron momentum component along the plasma density gradient, 
result in an increasingly strong thermal anisotropy, which may exaggerate the strength 
of the TAWI.

We examine here with a PIC simulation the expansion of an initially unmagnetized circular 
plasma cloud. The electron momentum can in this case not be subdivided into a component 
that is always parallel to the density gradient and into two components orthogonal to it, 
unless the electron moves exactly in the radial direction. This case is unlikely. Consecutive 
electron reflections by the electric field of the rarefaction wave will diminish different 
momentum components and the anisotropy between the temperature in the radial and azimuthal 
directions is not boosted by multiple reflections. The purpose of our simulation study is 
to assess if the slowdown of the electrons in the sheath potential alone is still strong 
enough to trigger the TAWI. 

Our results are as follows. The ambipolar electric field accelerates the protons in the 
simulations to a few percent of $c$. A TAWI develops within the density gradient of the 
rarefaction wave. This instability is similar to that observed in Ref. \cite{Thaury}. 
It drives radial current channels, which yield the growth of a transverse magnetic (TM) 
field component. The growth of the TM wave is followed by that of a secondary TE 
magnetowave. The rotational polarization of this TE wave is concluded from the 
simultaneous presence of in-plane magnetic vortices and out-of-plane magnetic fields 
that interfere with those driven by the TAWI. The combined magnetic field of both waves 
breaks the circular symmetry of the rarefaction wave and scatters rapidly the electrons 
within, thereby reducing their thermal anisotropy. The instability saturates in our 
simulation, when the total magnetic energy density is about 1\% of the electron thermal 
energy. 

The TE and TM magnetowaves develop in the same radial interval, which suggests their 
connection. We propose that the superposition of the electric field, which is 
induced by the magnetic field of the TAWI, with the radial electric field 
of the rarefaction wave triggers the growth of the TE wave by the misalignment of the 
plasma density gradient and the electric field vector. Such a system is unstable 
\cite{Saleem,Saleem2} and it results in an instability similar to a thermoelectric 
instability. 

The structure of this paper is as follows. Section 2 describes the PIC simulation code and 
the initial conditions. Our results are presented in section 3. Section 4 summarizes our
findings and it discusses how these could be verified experimentally.

\section{The simulation code and the initial conditions}
 
The 2D3V (resolves two spatial and three momentum dimensions) particle-in-cell simulation 
code we employ here is based on the virtual particle-mesh numerical scheme \cite{Eastwood}. 
Faraday's and Amp\`ere's laws 
\begin{eqnarray}
\nabla \times \mathbf{E} = -\frac{\partial \mathbf{B} }{\partial t} \label{Eq1} \\
\nabla \times \mathbf{B} = \mu_0 \mathbf{J} + \mu_0 \epsilon_0 \frac{\partial \mathbf{E}}{\partial t}
\label{Eq2}
\end{eqnarray}
are solved on a grid and evolve the electromagnetic fields in time, while $\nabla \cdot \mathbf{E}
=\rho / \epsilon_0$ and $\nabla \cdot \mathbf{B} = 0$ are fulfilled as constraints. The code solves 
all three field components as a function of $\mathbf{x}=(x,y)$. The relativistic force equation 
\begin{equation}
\frac{ d \mathbf{p}_i }{dt} = q_j \left ( \mathbf{E}[\mathbf{x}_i] + \mathbf{v}_i \times \mathbf{B}
[\mathbf{x}_i] \right ) \label{Eq3}
\end{equation} 
is used to update the momentum of the $i^{th}$ computational particle (CP) of species $j$, with 
charge $q_j$ and mass $m_j$, that is located at the position $\mathbf{x}_i$. The electric 
$\mathbf{E}$ and magnetic $\mathbf{B}$ fields are connected with the CPs as follows. The 
micro-current of each CP is interpolated to the grid. The sum of the interpolated micro-currents 
of all CPs gives the macroscopic current $\mathbf{J}$, which is used to update the electromagnetic 
fields through Eq. \ref{Eq2}. The new fields are interpolated back to the positions of the individual 
CPs to update their momentum in time through Eq. \ref{Eq3}. The position of each CP is updated with 
$d\mathbf{x}_i / dt =(v_{i,x},v_{i,y})$. A detailed review of the PIC method is given elsewhere 
\cite{Dawson}.

\begin{figure}
\begin{center}
\includegraphics[width=8cm]{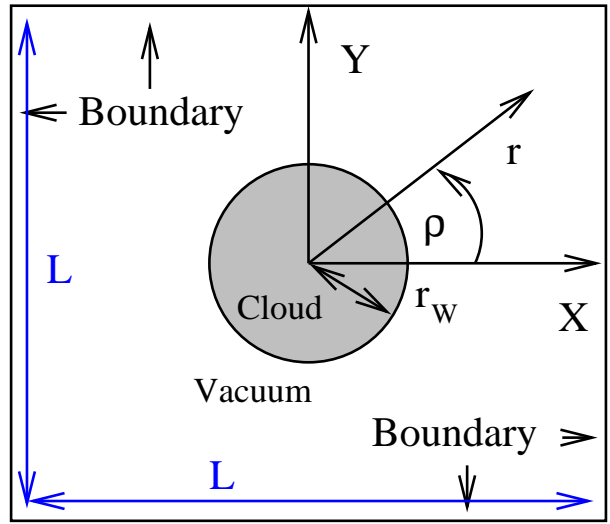}
\caption{The 2D simulation plane covers the interval $-L/2 < x,y < L/2$. A circular plasma cloud has 
its center at $x,y=0$ and the radius $r_W=L/5.4$. It is filled with electrons and protons with the 
density $n_0$. The boundary conditions are periodic for the fields and for the particles. We will 
in some cases exploit the circular symmetry and use the circular coordinates $(r,\rho)$ instead of 
$(x,y)$.}
\label{fig1}
\end{center}
\end{figure}

The plasma cloud with radius $r_W$ is surrounded by vacuum and it consists of electrons with the 
number density $n_0$ and protons with the same density, which are initially distributed uniformly 
across the cloud. Figure \ref{fig1} illustrates the setup.
The electron's relativistic Maxwellian velocity distribution has a temperature of 32 keV or a 
thermal speed $v_e = c/4$. The proton thermal speed in both simulations is $v_p = 3.1 \times 
10^4$ m$\textrm{s}^{-1}$ and corresponds to a thermal energy of 10 eV. The mean speeds of the 
electrons and of the protons vanish initially. The electrons and protons are approximated each 
by $\approx$ 2500 CPs per cell. The electron mass is $m_e$ and the proton mass $m_p = 1836 m_e$. 
The $\lambda_e = c / \omega_{p}$ is the initial electron skin depth and $\omega_p = {(n_0 e^2/m_e \epsilon_0)}^{1/2}$ is the electron plasma frequency. The time is expressed in 
units of $\omega_p^{-1}$. The time step is $\Delta_t = 0.022$. The simulation time is $T_F=880$.
The cloud radius $r_W = 14.2 \lambda_e$. An electron with the thermal speed $v_e$ will cross 
the cloud diameter $2r_W$ about 8 times during $t=T_F$, resulting in multiple bounces at the 
electrostatic sheath field. The simulation box with periodic boundary conditions in all 
directions uses $1700 \times 1700$ cells for a square domain with the side length $L=75.5 \lambda_e$.

\section{The simulation results}

We examine the plasma phase space distributions and the electromagnetic field distributions, 
the loss of electron thermal energy to the protons and the thermal anisotropy. We consider 
in more detail the simulation times $T_1 = 5.5$, $T_2 = 550$ and $T_3 = 750$.   This section 
concludes with a discussion of the time-evolution of the plasma quantities and effects due to 
the periodic boundary conditions. 

We transform the distributions from a representation in $(x,y)$-space into one in 
$(r,\rho)$-space, where $r = 0$ is the center of the cloud and of the simulation box. We exploit 
the approximate circular symmetry and average these quantities over the azimuth angle $\rho$. 
This average will become less accurate at late times, when $\mathbf{B}$ breaks this symmetry.
The electron number density $\tilde{n}_e (r) = \int n_e (r,\rho) r d\rho$. The phase 
space densities $f_{e,p}(r,\rho,v_r,v_\rho,v_z)$ are integrated as $\tilde{f}_{e,p}(r,v_r) 
= r^{-1} \int f_{e,p} \, r d\rho dv_\rho dv_z$, where $e$ and $p$ stand for electrons and 
protons. The phase space densities at all times are normalized to the maximum of the respective 
distribution at $t=0$. The azimuthally averaged radial thermal energy $K_r (r) = K_0^{-1} 
\int_\rho r d\rho \int f_e v_r^2 d\mathbf{v}$ and azimuthal thermal energy $K_\rho (r)
 = K_0^{-1}\int r d\rho \int f_e v_\rho^2 d\mathbf{v}$ define the thermal anisotropy $A(r) 
= K_r (r) / K_\rho (r)$. The initial electron thermal energy per degree of freedom is $K_0$. 
The in-plane electric field $E_p = |E_x + iE_y|$ and the orthogonal $E_z$ field are expressed 
in units of $\omega_p c m_e / e$, while the magnetic field components $B_p = |B_x + i B_y|$ 
and $B_z$ are expressed in units of $\omega_p m_e / e$. The fields are azimuthally averaged as 
$\langle F^2 \rangle = {(2\pi r)}^{-1} \int_\rho F^2 r d\rho$, where $F$ serves as a place 
holder for $E_p, E_z, B_p$ and $B_z$.
 
\subsection{Early time T1: Proton acceleration}

\begin{figure}
\includegraphics[width=8cm]{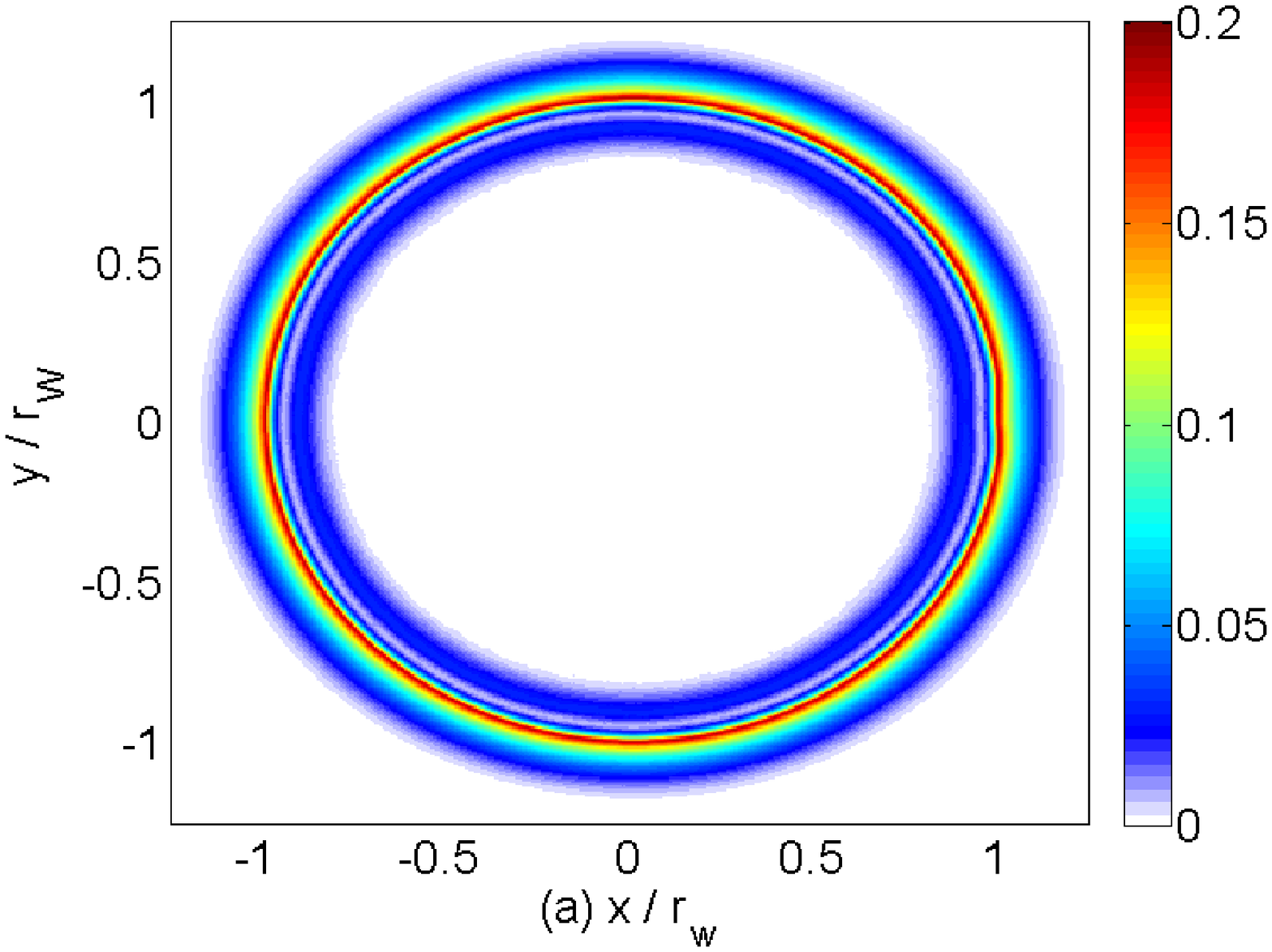}
\includegraphics[width=8cm]{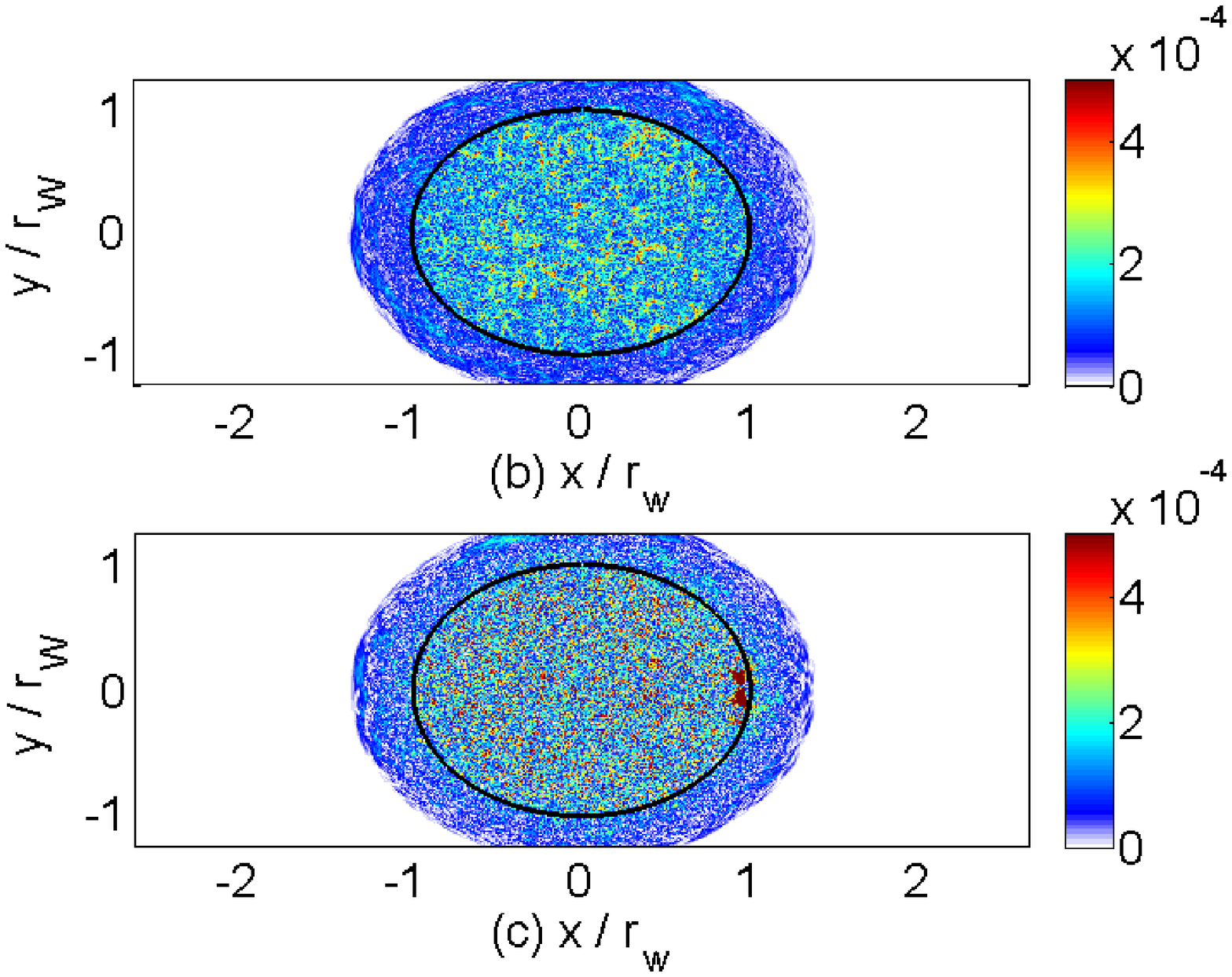}
\caption{(Color online) The field distributions at the time $T_1$: (a) shows the modulus 
of the in-plane electric field $E_p$. (b) displays the in-plane magnetic field $B_p$ and 
(c) the orthogonal field $|B_z|$. A strong electric field points radially outwards and 
outlines the cloud's perimeter. The magnetic fields are at noise levels. Overplotted is 
a black circle with $r=r_W$.}\label{fig2}
\end{figure}

Their mobility implies that some electrons can escape from the cloud and build up a 
negatively charged sheath outside of the now positively charged cloud. An ambipolar 
electrostatic field develops that points along the surface normal. Its modulus is displayed in 
Fig. \ref{fig2} for the time $T_1$.
The electric field outlines the cloud radius and has a thickness comparable to $\lambda_e$. 
The magnetic $B_p$ and $B_z$ components show noise that is strongest for 
$r < r_W$ and weaker in the rarefaction wave. The noise originates from statistical 
fluctuations of the current density. The initial plasma evolution is dominated by the 
electrostatic fields because their amplitude is much larger than the amplitude of the magnetic 
fields and because the latter are incoherent.

\begin{figure}
\begin{center}
\includegraphics[width=8cm]{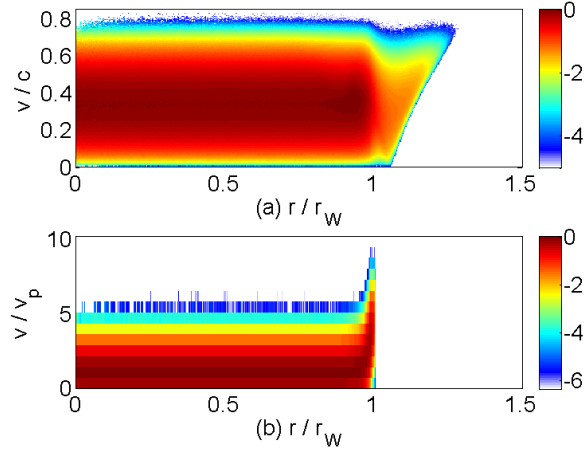}
\caption{(Color online) The azimuthally averaged radial phase space density at the time $T_1$ normalized to 
its initial value: Panels (a) and (b) show $\tilde{f}_e$ and $\tilde{f}_p$, 
respectively. The color scale is 10-logarithmic and $r_W$ is the initial cloud radius.
\label{fig3}}
\end{center}
\end{figure}

The protons are accelerated radially outwards by the electric field, which confines at the 
same time the electrons. This is confirmed by Fig. \ref{fig3}. The electrons lose speed in 
the radial direction and their density decreases drastically as they cross the strong 
in-plane electric field $E_p$ at $r=r_W$. This field is not strong enough to hold back all 
electrons and some escape into the vacuum. The fastest electrons have expanded to a radius 
$r \approx 1.3 r_W$, while the proton expansion is still negligible. The protons with $r\approx r_W$ are 
accelerated by $E_p$ and some reach already at the time $T_1$ a peak speed that exceeds 
their initial thermal speed $v_p$ by an order of magnitude. 
The protons with $r<r_w$ still show their initial velocity distribution. Their low thermal speed implies, that they could not yet 
interact with the surface electric field. 

\subsection{Intermediate time T2: The TAWI}

\begin{figure}
\begin{center}
\includegraphics[width=8cm]{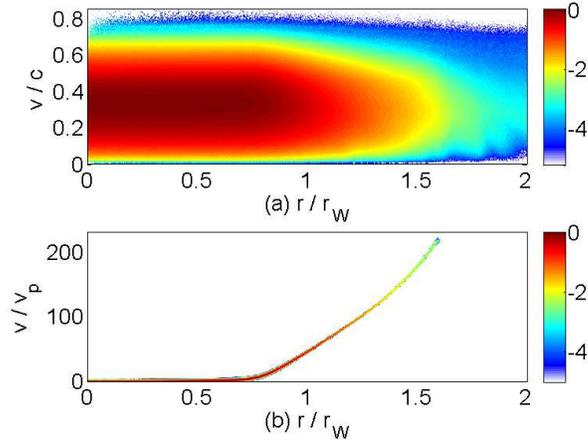}
\caption{(Color online) The azimuthally averaged radial phase space density at the time $T_2$ normalized to its 
initial value: Panels (a) and (b) show $\tilde{f}_e$ and $\tilde{f}_p$, respectively. 
The color scale is 10-logarithmic and $r_W$ is the initial cloud radius.\label{fig4}}
\end{center}
\end{figure}

Figure \ref{fig4} shows the particle distributions at the time $T_2$. The protons have been 
accelerated by the ambipolar electric field to about $200 v_p$ or to 2\% of the speed of light 
$c$. The protons form a narrow beam with a speed, which grows approximately linearly with $r$ 
for $r>0.7r_W$, and a density that decreases with an increasing radius. This high speed has been 
reached by the protons, because there is no shock that limits their acceleration. A shock would 
form, if the rarefaction wave would expand into an ambient plasma. 
The electron distribution has been altered significantly at this time. The peak speed of the 
electron's bulk population in the cloud's core, which has shrunk to $r<0.7r_W$ at this time, 
has been decreased compared to that in Fig. \ref{fig3}(a), which is indicative of an energy 
loss. This energy loss can be attributed to the electron's inelastic reflection by the ambipolar 
electric field of the expanding rarefaction wave. The electrons fill up the entire simulation box, 
although their phase space density close to the boundary is 4 orders of magnitude less than the 
peak density within $r<0.7r_W$. 

\begin{figure}
\includegraphics[width=8cm]{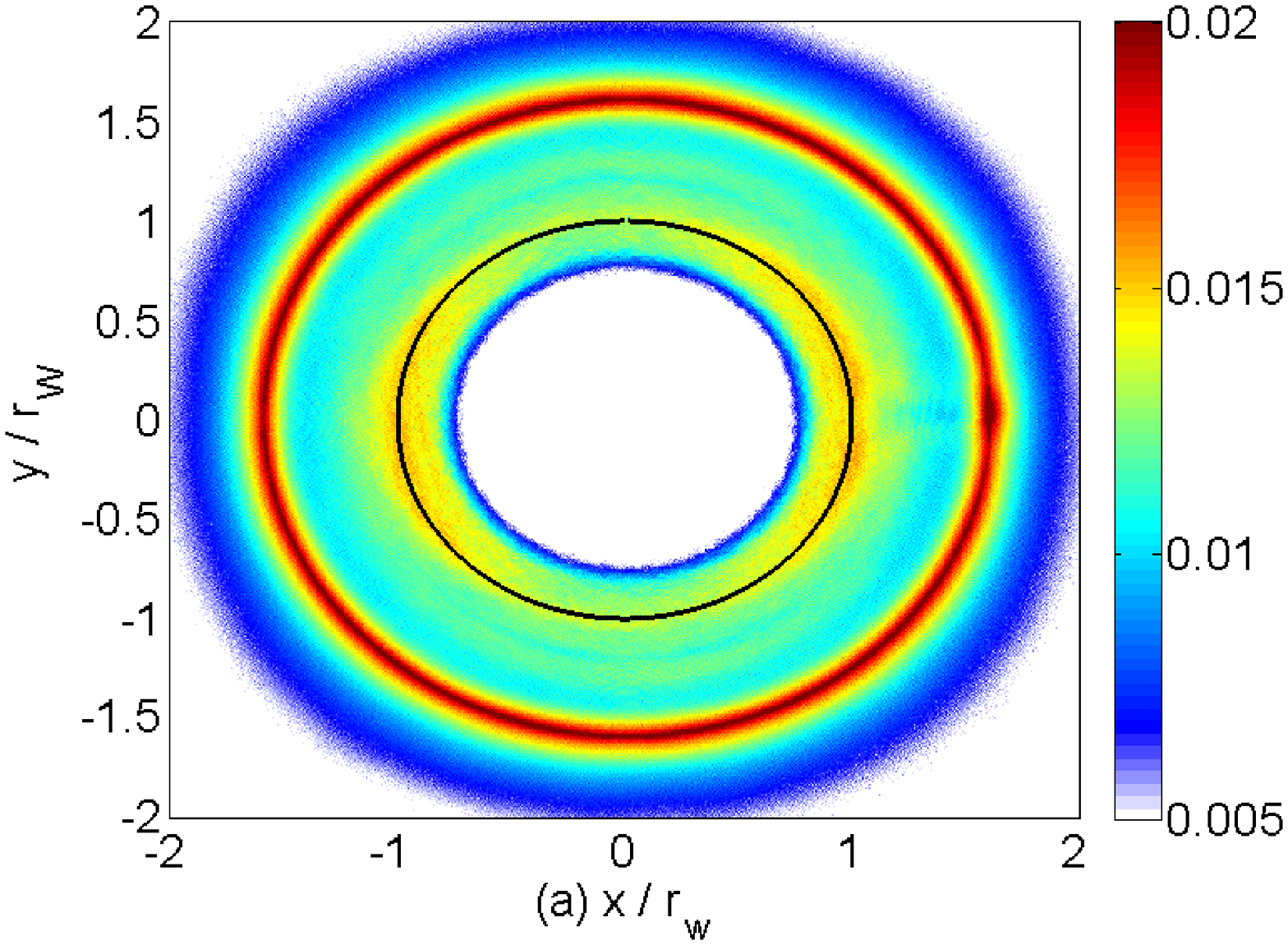}
\includegraphics[width=8cm]{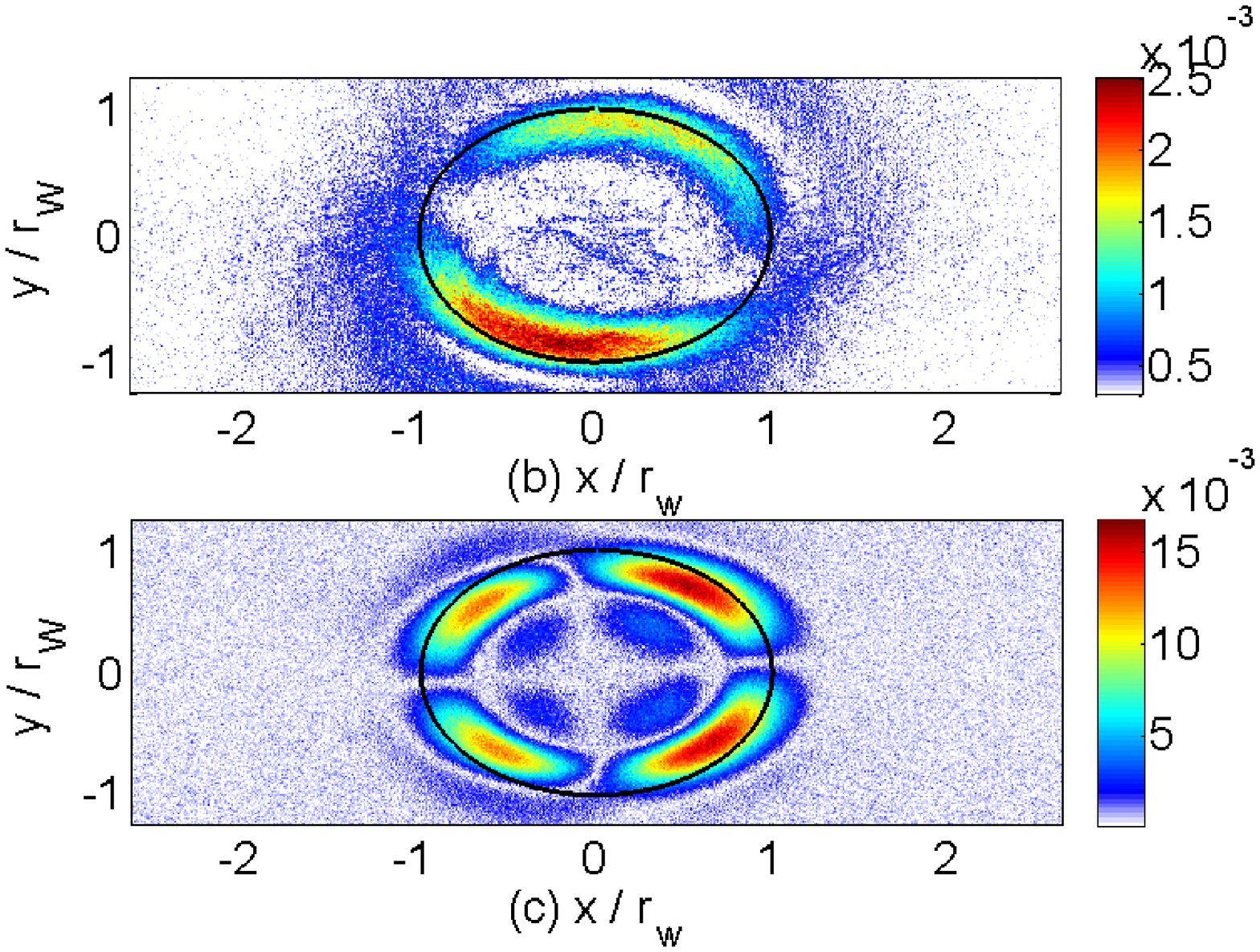}
\caption{(Color online) The field distributions at t=$T_2$: Panel (a) shows the modulus 
of the in-plane electric field $E_p$ and (b) shows that of the in-plane magnetic field 
$B_p$. Panel (c) shows the modulus $|B_z|$. The black circles with radius $r_W$ outline 
the initial cloud perimeter.\label{fig5}}
\end{figure}

The erosion of the cloud implies, that the initial distribution of $E_p$ in Fig. \ref{fig2}(a) 
could not have been upheld. This is confirmed by Fig. \ref{fig5}, which shows the $E_p$, $B_p$ 
and the modulus of $B_z$ at the time $T_2$.
The electric field maximum is located at $r\approx 1.6r_W$, which is at the tip of the 
expanding protons in Fig. \ref{fig4}(b). The electric field distribution expands to lower 
$r$. It thus continues to accelerate the protons and sustains the increase of the proton 
mean speed with the radius. The electric fields at $r\approx r_W$ are a consequence of the 
plasma density gradient in this interval and of the higher electron mobility, which implies 
a space charge. Figure \ref{fig5}(a) reveals, that the electric field in the interval 
$r\approx r_W$ does not show an exact circular symmetry. The amplitude is a function of 
$\rho$.

\begin{figure}
\begin{center}
\includegraphics[width=7.7cm]{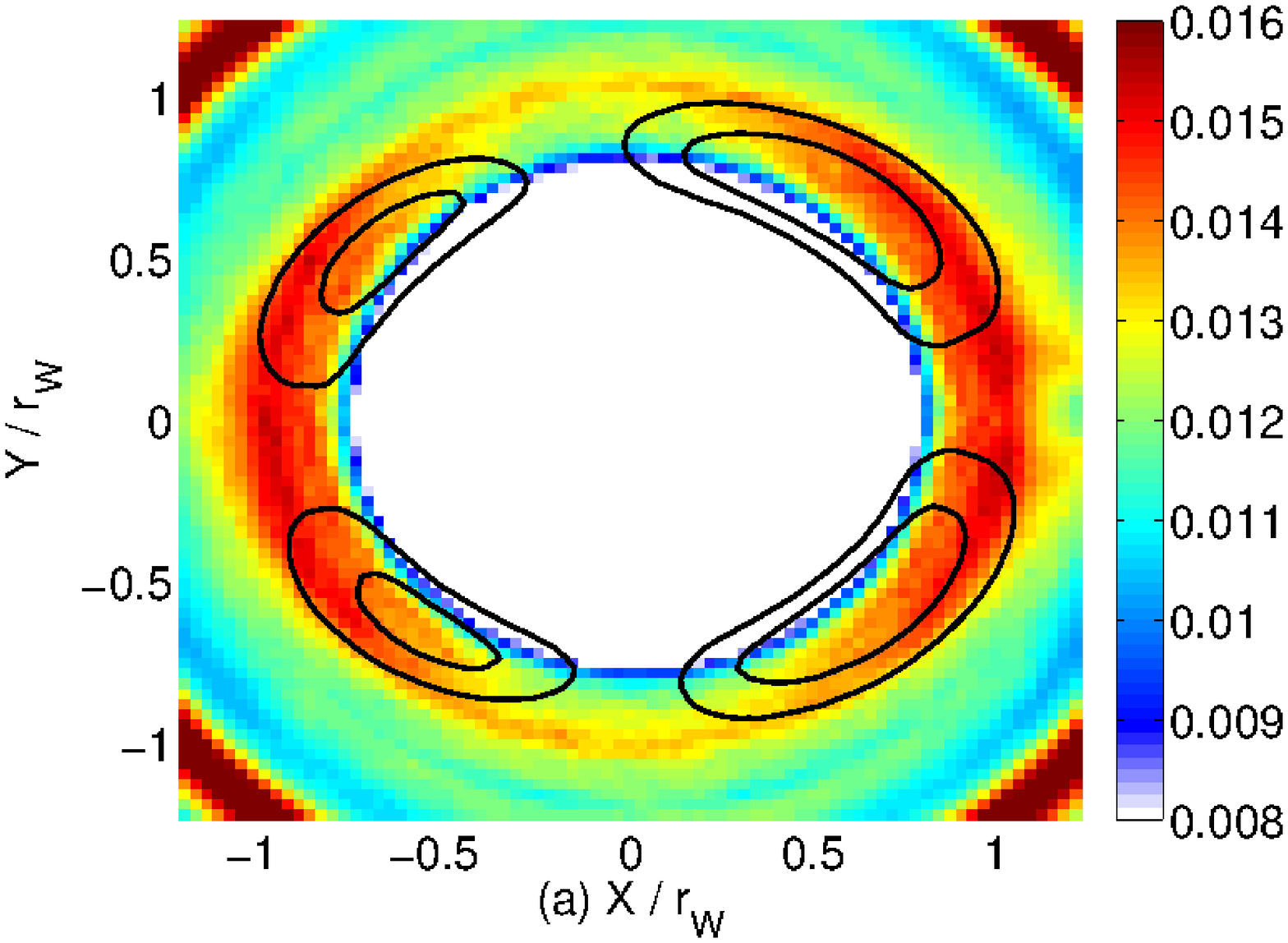}
\includegraphics[width=7.7cm]{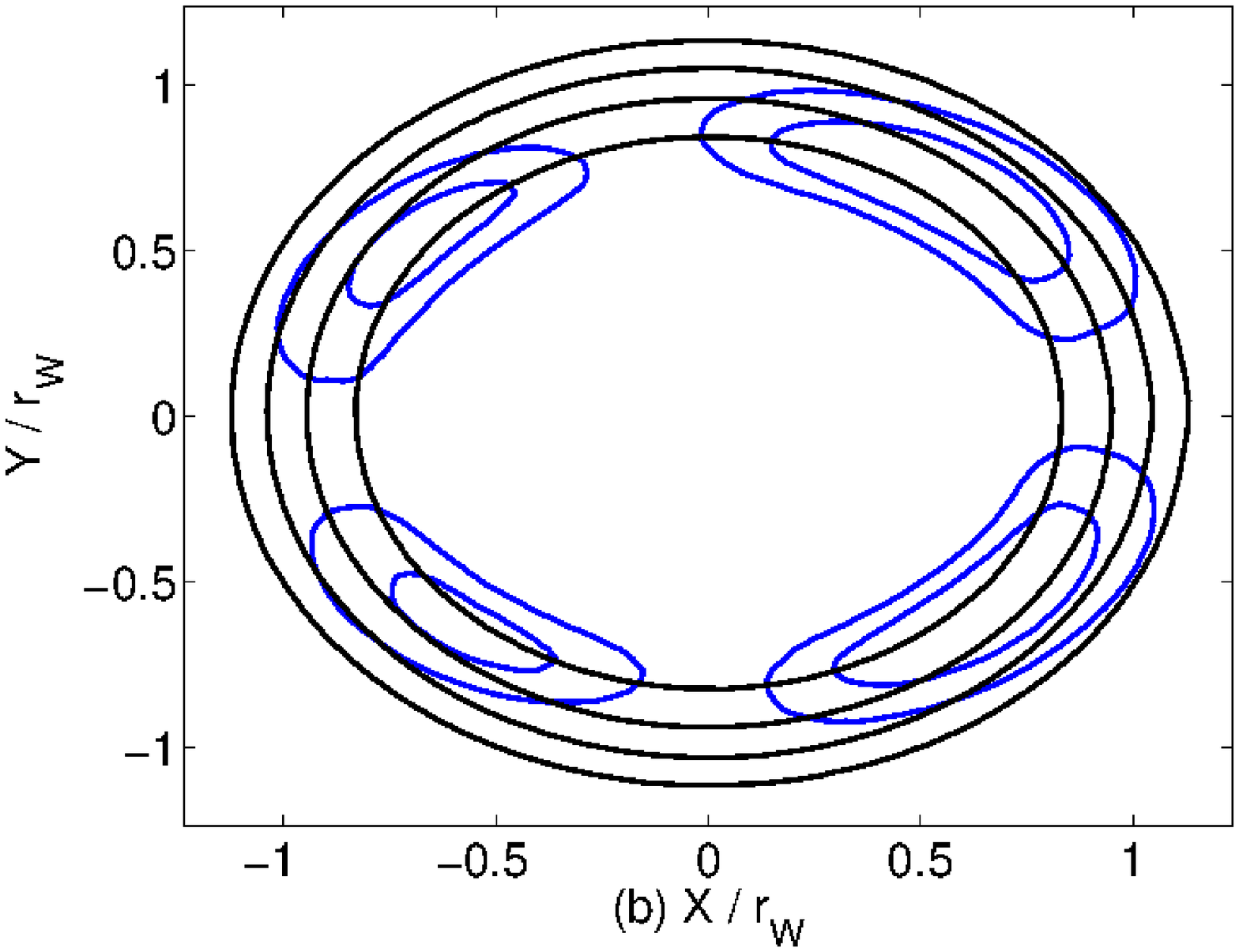}
\caption{Panel (a) shows the modulus of the in-plane electric field $E_p (x,y)$. The 
contour lines correspond to the values -0.01, -0.005, 0.005 and 0.01 of the $B_z$ field.
The $B_z$ field has maxima in the upper right and lower left quadrant and minima
in the other two quadrants. Panel (b) compares the same contour lines of $B_z$ with 
the contour lines 2.7, 2.9, 3.1 and 3.3 of the 10-logarithmic electron number density. The 
density increases with decreasing radius. The number density is expressed in computational 
particles per cell and $\log_{10} n_0 = 3.4$.}\label{fig6}
\end{center}
\end{figure}

This break of the circular symmetry and its origin become evident in Fig. \ref{fig6}, 
which shows the amplitude modulus of $E_p$ and the electron density distribution in 
the relevant spatial interval and compares both quantities to $B_z$. The amplitude of the 
azimuthal oscillation of $E_p$ at $r\approx r_W$ is about 10\% of the mean value. 
Figure \ref{fig6}(a) suggests a correlation of $E_p$ with $B_z$ for two reasons. Firstly the 
azimuthal oscillation of $E_p$ and the strong $B_z$ are observed at approximately the same 
radius. Secondly the $E_p$ field shows two amplitude minima and two maxima, which are shifted 
by 45$^\circ$ relative to those of $B_z$. We find one maximum of $E_p$ at $x \approx r_W$ and 
$y \approx 0$ ($\rho \approx 0^\circ$), while one maximum of $B_z$ is located at $x, y \approx 
r_W / \sqrt{2}$ ($\rho \approx 45^\circ$).  

The circular symmetry of $E_p$ can, in principle, be broken by an electrostatic drift wave 
along the azimuthal direction. They could be generated by the lower-hybrid drift instability 
\cite{Lapenta1,Lapenta2}, which is caused by gradients in the plasma density and in the magnetic 
field amplitude. A second possibility is the Simon-Hoh instability that develops, if a radial 
electrostatic field and a spatially uniform orthogonal magnetic field are present 
\cite{SimonHoh,Dawson2}. The electron $\mathbf{E} \times \mathbf{B}$ drift destabilizes the 
plasma. The electric field driving the Simon-Hoh instability would be the $E_p$ and the 
orthogonal magnetic field would be $B_z$. However, the characteristic wave lengths of the waves 
driven by both instabilities would be much less than that of the modulation of $E_p$ observed 
in Fig. \ref{fig6}. The lower hybrid instability destabilizes wavelengths, which are shorter 
than the characteristic length scale of the magnetic field gradient and the Simon-Hoh instability 
requires that $B_z$ be uniform over a scale much larger than the wave length of the unstable wave. 
Figure \ref{fig6}(b) furthermore demonstrates that the electron density distribution is still 
circularly symmetric. The azimuthal oscillation of $E_p$ is thus not correlated with a charge 
density modulation and it is probably not an electrostatic wave.

We can understand the modulation of $E_p$, if we take into account the 
in-plane current $(J_x,J_y)$ and electric field $(E_x,E_y)$ that sustain the localized 
$B_z$. We neglect the weak $B_p$ in Fig. \ref{fig5}(b). Amp\'ere's law (Eq. \ref{Eq2}) 
connects $B_z$ with the in-plane electric field and the in-plane current. Consider the right 
part of the simulation box $x>0$ in Fig. \ref{fig6}(a). The contour lines with $y>0$ denote a 
maximum of $B_z$ and those with $y<0$ a minimum. At $x\approx r_W$ and $y \approx 0$ we find
with the right hand rule that the electric field induced by $\nabla \times (0,0,B_z)$ gives $\partial_t
E_x > 0$ and $\partial_t E_x \gg |\partial_t E_y|$. The electric field amplitude should grow to a sizeable amplitude before a balance between the in-plane current and $B_z$ cancels any further growth of $E_p$. The superposition of 
the ambipolar electric field from the rarefaction wave and the practically parallel induced 
$E_x$ at $x\approx r_W$ and $y\approx 0$ imply that $E_p$ is strong. The induced electric field will also amplify the ambipolar 
electric field at $x\approx -r_W$ and $y\approx 0$, because both fields reverse their sign. 
The induced electric field weakens the ambipolar electric field at $x\approx 0$ and $|y| \approx 
r_W$. A magnetically induced electric field is not tied to charge density modulations, which would 
explain why the circular symmetry is maintained for the electron density in Fig. \ref{fig6}(b).

\begin{figure}
\begin{center}
\includegraphics[width=8cm]{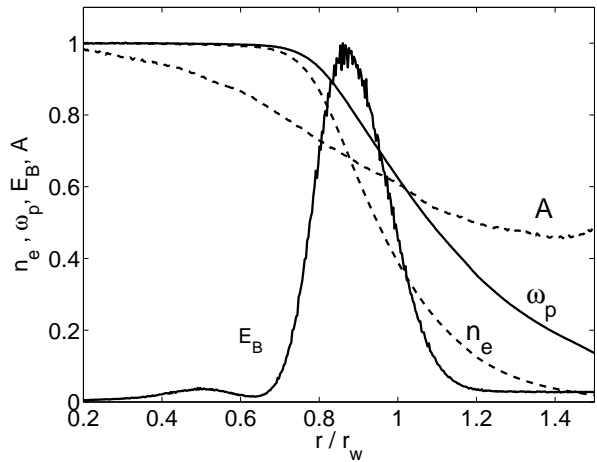}
\caption{The azimuthally averaged plasma configuration at t=$T_2$: The (dashed) curve 
$n_e$ corresponds to the radially averaged electron number density $\tilde{n}_e / r$, 
which is normalized to $n_0$. The curve $\omega_p$ equals ${(\tilde{n}_e / n_0 r)}^{0.5}$ 
and denotes the local plasma frequency. The curve $E_B$ is the total magnetic energy 
density $r \langle B^2_p \rangle + r \langle B^2_z \rangle$. The thermal anisotropy $A = K_r 
(r) / K_\rho (r)$ is the second dashed curve.}\label{fig7}
\end{center}
\end{figure}

We turn now to the identification of the source mechanism of the magnetic $B_z$ component 
in Fig. \ref{fig5}(c). Figure \ref{fig7} compares the azimuthally averaged energy density 
$r (\langle B_p^2 \rangle + \langle B_z^2 \rangle)$ of the magnetic field, the thermal 
anisotropy $A(r) = K_r (r) / K_\rho (r)$ and the azimuthally averaged electron number 
density $\tilde{n}_e / r$. The division of $\tilde{n}_e$ by $r$ ensures that a uniform 
density within the cloud results in a constant value of $\tilde{n}_e / r$ and the 
multiplication of the azimuthally averaged energy density by $r$ gives us a curve that is 
proportional to the magnetic energy. These quantities are computed at the time $T_2$.

The onset of the rarefaction wave is found at $r \approx 0.6 r_W$ in Fig. \ref{fig7}, where 
the electron density and plasma frequency start to decrease. A sidelobe of the magnetic 
energy $E_B$ is found within this radius, but most energy is concentrated within $0.7 < r / 
r_W < 1.1$. The thermal anisotropy curve is a steadily decreasing function up to $r \approx 
1.4 r_W$ and it is below 1 everywhere. The mean kinetic energy of the electrons in the radial 
direction is thus lower than their azimuthal one, which we attribute to the radial proton 
acceleration. The TAWI destabilizes waves with wave vectors along the cool direction. Figure 
\ref{fig5}(c) shows that $B_z$ oscillates fastest in the radial direction. One oscillation 
is visible with the wavelength $\approx 0.8r_W$. The magnetic $B_z$ component performs two 
oscillations in the azimuthal direction, each with a wavelength $\approx \pi r_W$. A long 
oscillation orthogonal to the density gradient has also been observed in simulations of a 
planar rarefaction wave \cite{Thaury}, which is probably a consequence of the density 
gradient that has not been taken into account by the existing analytic studies. 

The simulation data suggests that the large anisotropy observed in Fig. \ref{fig7} triggers the 
TAWI. The $E_B$ peaks at $r \approx 0.85 r_W$, where $A = 0.7$. One may think that $E_B$ should 
peak in the interval, where $A$ is lowest. However, both the growth rate of the TAWI in units 
of $s^{-1}$ and the magnetic field amplitude the plasma can support increase with $\omega_p$. 
The latter can be concluded from the fact, that we can normalize the Vlasov-Maxwell set of 
equations with the help of $\omega_p$. A higher $\omega_p$ results in a higher $\mathbf{B}$ for 
the same plasma configuration. The magnetic amplitude decreases to noise levels at about $r/r_W 
\approx 1.5$. It is thus confined to within the rarefaction wave and it does not extend into the 
surrounding electron plasma at this time. 

\subsection{Final time T3: Secondary instability}

The electrons have spread out in significant numbers up to $r\approx 2r_W$ at 
the time $T_3$. 
\begin{figure}
\begin{center}
\includegraphics[width=8cm]{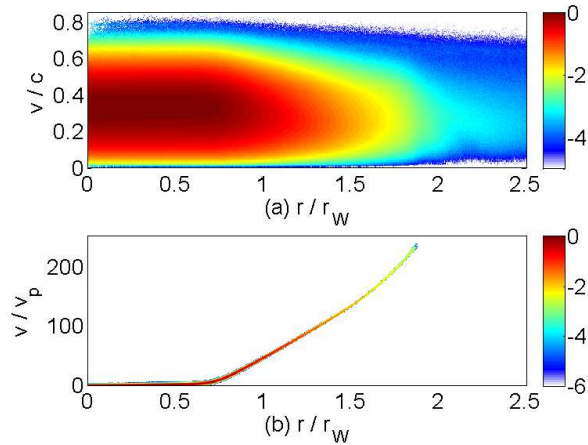}
\caption{(Color online) The azimuthally averaged radial phase space density at the time $T_3$ 
normalized to its initial value: Panels (a) and (b) show $\tilde{f}_e$ and $\tilde{f}_p$, 
respectively. The color scale is 10-logarithmic and $r_W$ is the initial cloud radius.}\label{fig8}
\end{center}
\end{figure}
The radial electric field has accelerated the protons to about 260$v_p$ or 2.6\% of c at the 
time $T_3$, as we can see from Fig. \ref{fig8}(b). The protons form a cool beam with 
one well defined mean speed for any $r\le 1.8r_W$. This compact distribution justifies its 
approximation by a single cold fluid \cite{Grismayer}. The mean speed of the protons increases 
linearly within $0.7 < r/r_W < 1.6$. 

\begin{figure}
\includegraphics[width=8cm]{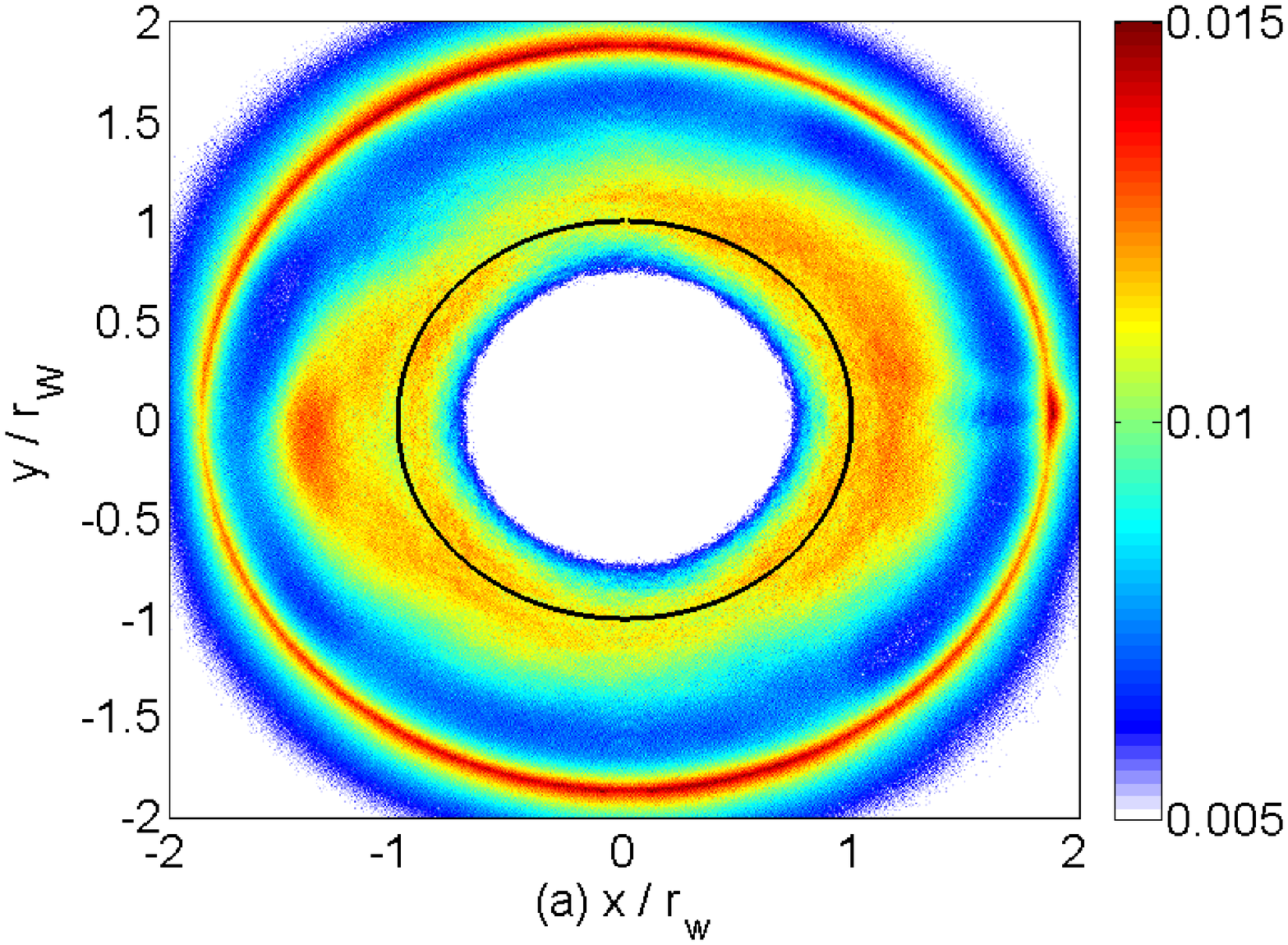}
\includegraphics[width=8cm]{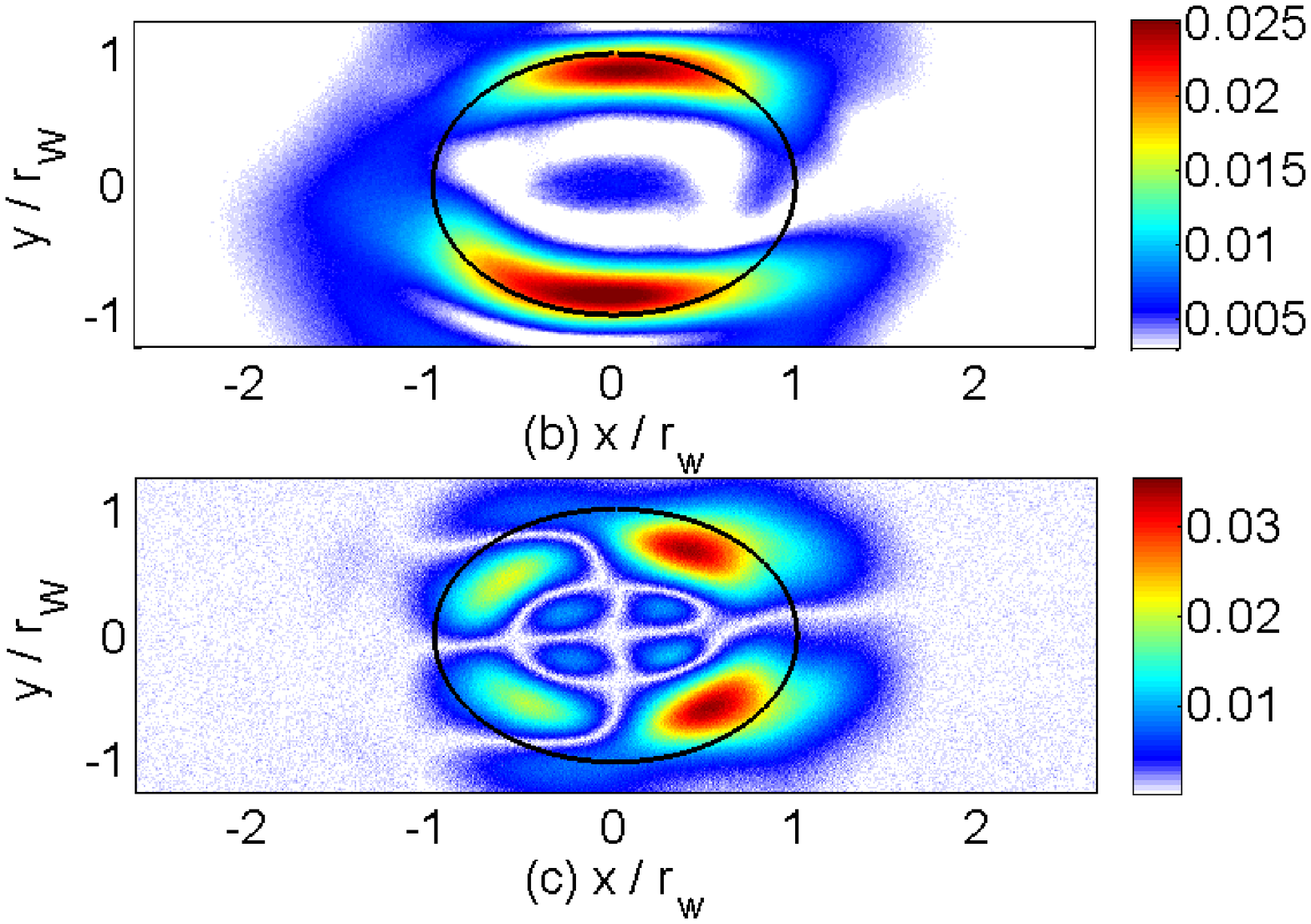}
\caption{(Color online) The field distributions at t=$T_3$: Panel (a) shows the in-plane 
electric field $E_p$ and (b) shows the in-plane magnetic field $B_p$. Panel (c) shows the 
modulus $|B_z|$. The black circles with radius $r_W$ outline the initial cloud perimeter.
}\label{fig9}
\end{figure}

Figure \ref{fig9} shows $E_p$, $B_p$ and $|B_z|$ at the time $T_3$. 
The in-plane electric field has weakened while the magnetic fields have grown further. 
The plasma density gradient and the ambipolar electric field in a rarefaction wave 
decrease with time. The influence on the particle dynamics of the magnetic 
field and its induced electric field thus become more important. The circular symmetry 
of $E_p$ imposed by the rarefaction wave is no longer dominant in Fig. \ref{fig9}(a). 
The amplitude of the in-plane $B_p$ in Fig. \ref{fig9}(b) has increased by an order of 
magnitude compared to that in Fig. \ref{fig5}(b) and it is now comparable to the amplitude 
of $B_z$. The distribution of $|B_z|$ in Fig. \ref{fig9}(c) does no longer show four 
almost equally important peaks as it did in Fig. \ref{fig5}(c). 

We would expect such a pattern, if the wave responsible for the $B_p$ is driven by a separate 
secondary instability and if it results in a circularly or elliptically polarized magnetowave. An instability is discussed in the next subsection that can generate such a wave.
The $B_p$ field oscillates much faster in the radial than in the azimuthal direction in Fig. 
\ref{fig9}(b) and its wave vector is thus practically aligned with the radius vector. We call 
this magnetowave \textit{wave 2}. An elliptically polarized wave will have a $B_z$ component 
that is phase shifted along $r$ with respect to its $B_p$ component. The $B_z$ component of 
wave 2 would interfere with the $B_z$ component of the wave that is driven by the TAWI (wave 1). 
Wave 2 performs one oscillation in the azimuthal direction according to Fig. \ref{fig9} and we 
may assume that the same is true for its $B_z$ component. The $B_z$ field of wave 1 oscillates 
twice along $\rho$. So we expect that wave 2 interferes constructively with two half-oscillations  
of wave 1 and that it interferes destructively with the other two. Interference between both 
waves could thus explain the pattern observed in Fig. \ref{fig9}(b). 

\subsection{Time evolution}

Figure \ref{fig7} has demonstrated that the magnetic field energy density is increased in the 
radial interval, in which we find a strong thermal anisotropy in the electron distribution. The anisotropy 
is caused by the loss of electron thermal energy to the radially accelerating protons. The
supplementary movie 1 animates in time this energy exchange. It shows the electron phase space distribution in (a) and the proton phase space distribution in (b), 
which are displayed in the form of snapshots and discussed in Figs. \ref{fig3}, \ref{fig4} and 
\ref{fig8}. Movie 1 confirms that electrons, which cross the boundaries on one side, re-enter 
on the opposite side, as we expect from the periodic boundary conditions of the simulation. The 
fastest electrons that escaped from the plasma cloud reach the boundaries at $t\approx 50$ and 
they encounter the electrons that crossed the boundary from the other direction. Note that
movie 1 (a) shows the velocity modulus and does thus not reveal that both electron populations
move into opposite directions. The electron two-stream configuration can result in instabilities. 
However, the two-stream configuration is rapidly destroyed by the arrival of the slower electrons 
at the boundaries. 

The supplementary movie 2 demonstrates that the life-time of the two-stream configuration is
too short and the electron density too low to drive detectable instabilities at $t\approx 50$. 
It animates in time the 10-logarithmic electron density in the simulation box, which is averaged 
over blocks of $4\times 4$ cells and given in particles per cell. An average density of less 
than one computational particle per cell is found close to the boundaries. The local plasma 
frequency is  $\approx \omega_p / 100$, which implies a low growth rate for any beam instability 
\cite{Review,Bret2}. Movie 2 reveals faint electron structures at a time $t\approx 200$ that flow 
parallel to the x-axis and y-axis. They are probably caused by an instability between the 
electrons of the plasma cloud and the electrons, which return from the boundary and are 
accelerated by the ambipolar electric field when they re-enter the cloud \cite{Shock1}.

\begin{figure}
\begin{center}
\includegraphics[width=7.7cm]{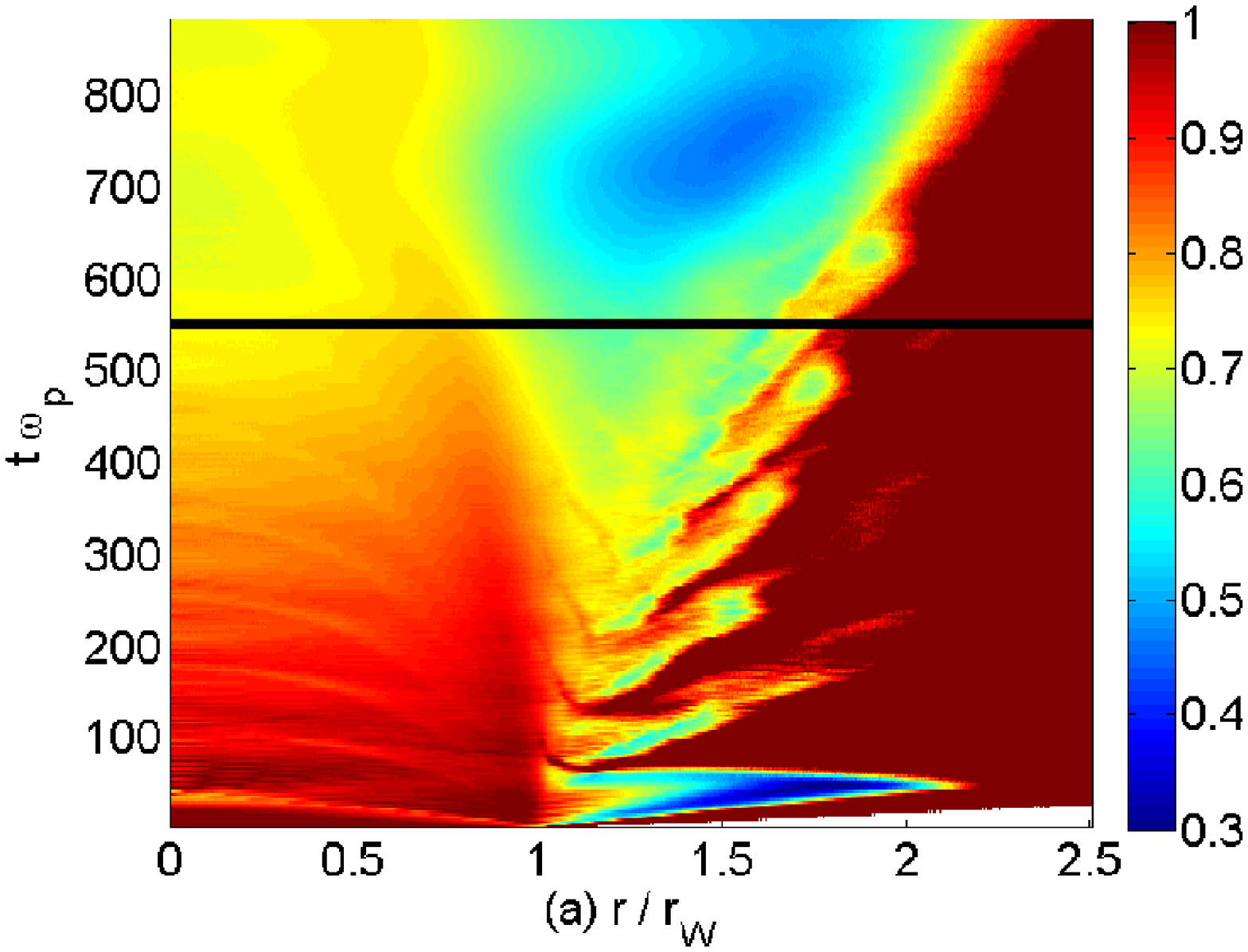}
\includegraphics[width=7.7cm]{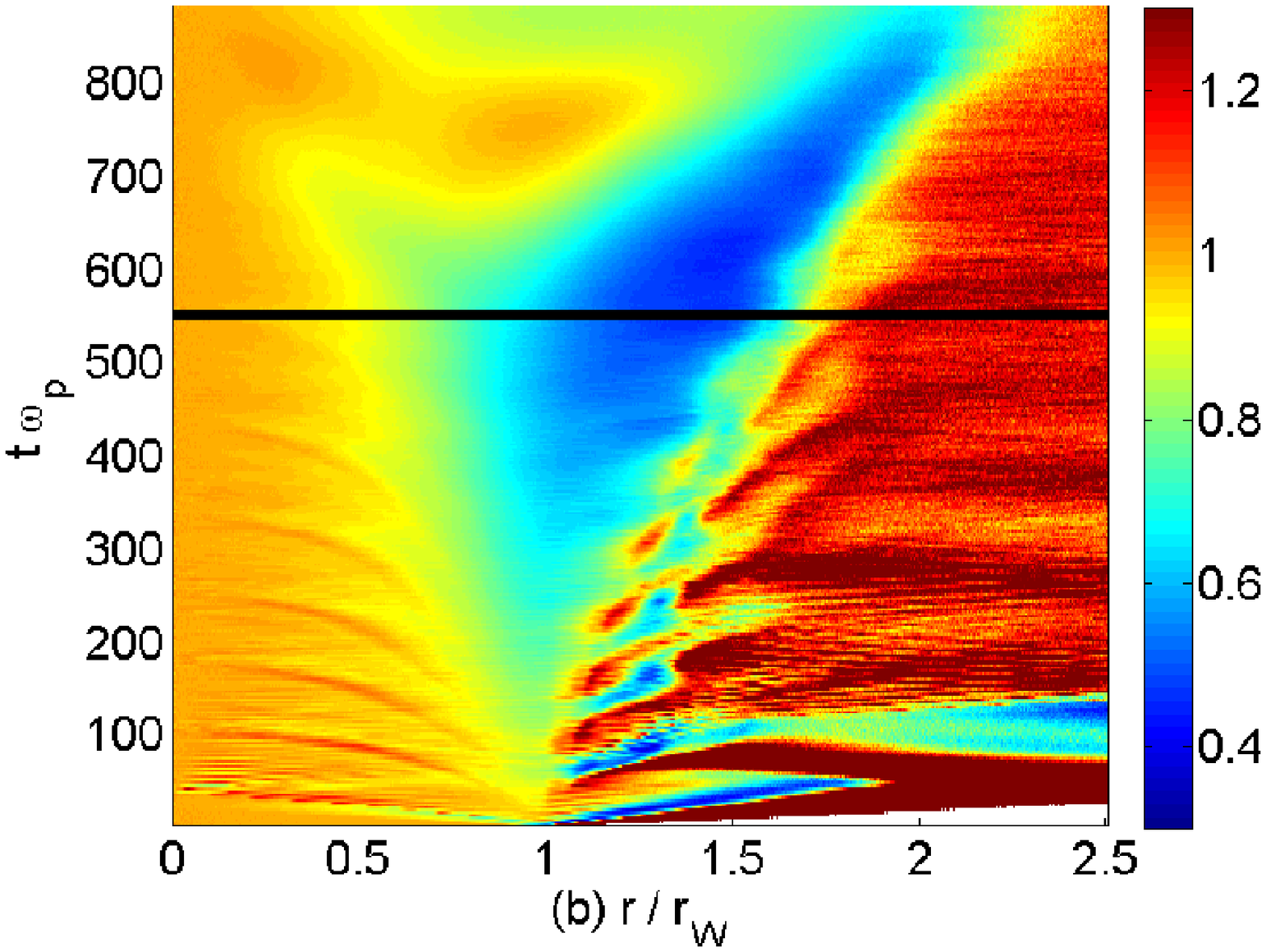}
\caption{(Color online) Panel (a) displays the time evolution of the electron thermal energy 
$n_0 (K_r (r) + K_\rho (r)) / 2 K_0 \tilde{n}_e$ and panel (b) the thermal anisotropy $A(r)=
K_r (r)/ K_\rho (r)$. The colour scale is linear. The horizontal black line corresponds 
to the time $T_2$}\label{fig10}
\end{center}
\end{figure}

Figure \ref{fig10} quantifies both the energy loss of the electrons to the accelerating 
protons and it follows the thermal anisotropy in time. Figure \ref{fig10}(a) is a measure 
for the average kinetic energy per electron, which corresponds to their velocity in the 
simulation plane. This energy is obtained by summing $K_r (r)$ and $K_\rho (r)$ and by 
dividing the result by $2K_0 \tilde{n}_e / n_0$. $K_0$ is the initial thermal energy 
contributed by each velocity component and the division by $\tilde{n}_e/n_0$ gives us a 
measure of the kinetic energy per electron. The thermal anisotropy $A(r,t)$ is displayed 
in Fig. \ref{fig10}(b).

The electrons within the cloud lose about 30\% of their kinetic energy to the accelerating 
protons. The electron kinetic energy is apparently increased outside of the rarefaction wave. 
This is a selection process, which can be seen in the movie 1. The bulk of the slow electrons 
is contained by the rarefaction wave, while the most energetic electrons, which have a kinetic 
energy that far exceeds the potential energy of the rarefaction wave, can escape. 

The thermal anisotropy is negligible for the radial interval without a plasma density gradient,
e.g. for $r<0.5r_W$ and $t=T_2$. $A(r)$ reveals that the electrons in the interval with the
proton density gradient lose more energy in the radial than in the azimuthal direction. The 
radial electron energy is less than half of the azimuthal energy over an extended radial 
interval. The interval with a value of $A(r)<0.6$ is largest at the time $T_2$ in Fig. 
\ref{fig10}(b), when it covers $0.9 < r / r_W < 1.3$. The $B_z$ saturates at around this time.
The TAWI starts to grow at $t \approx 200$, when the anisotropy is sufficiently strong. 

\begin{figure}
\includegraphics[width=8cm]{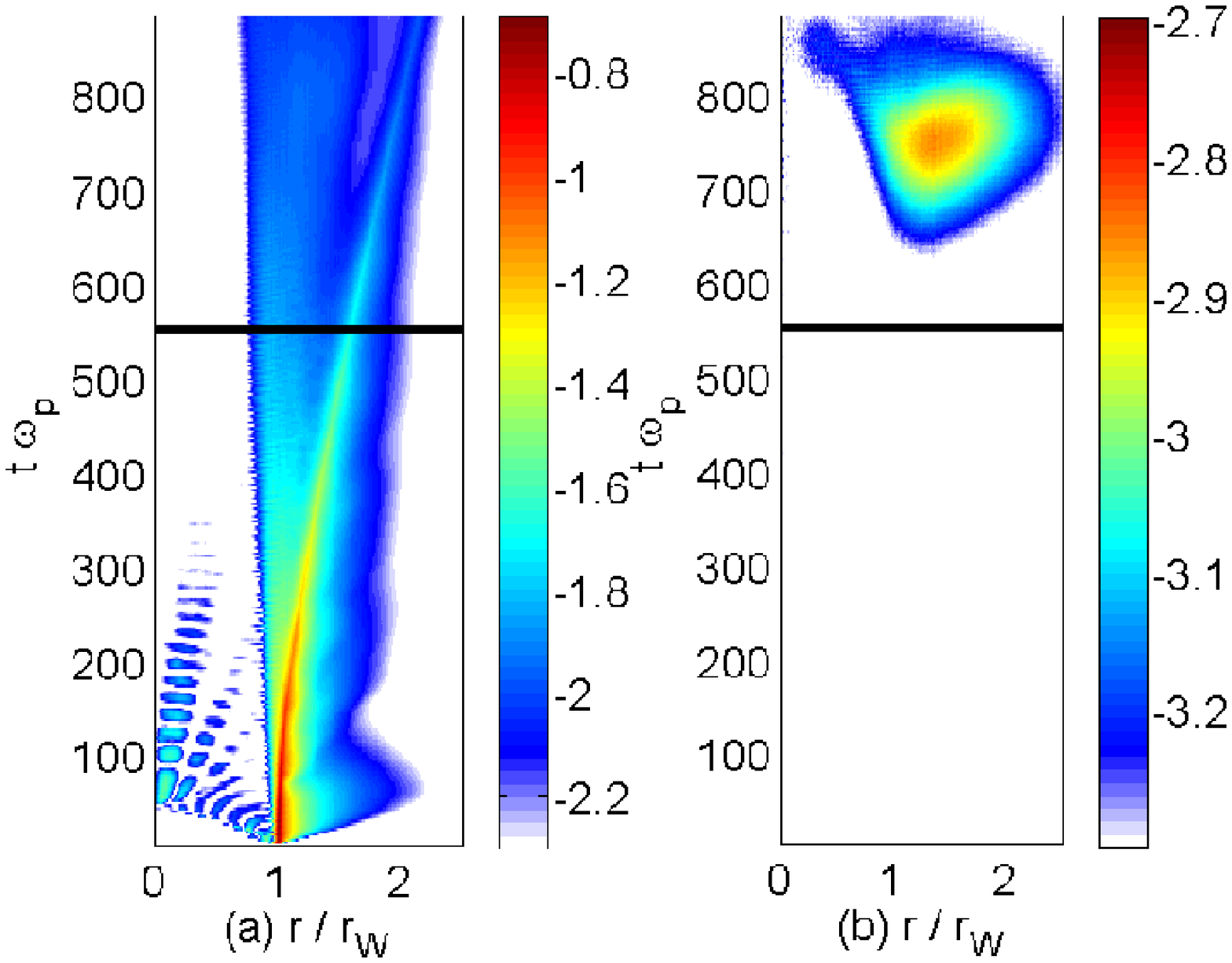}
\includegraphics[width=8cm]{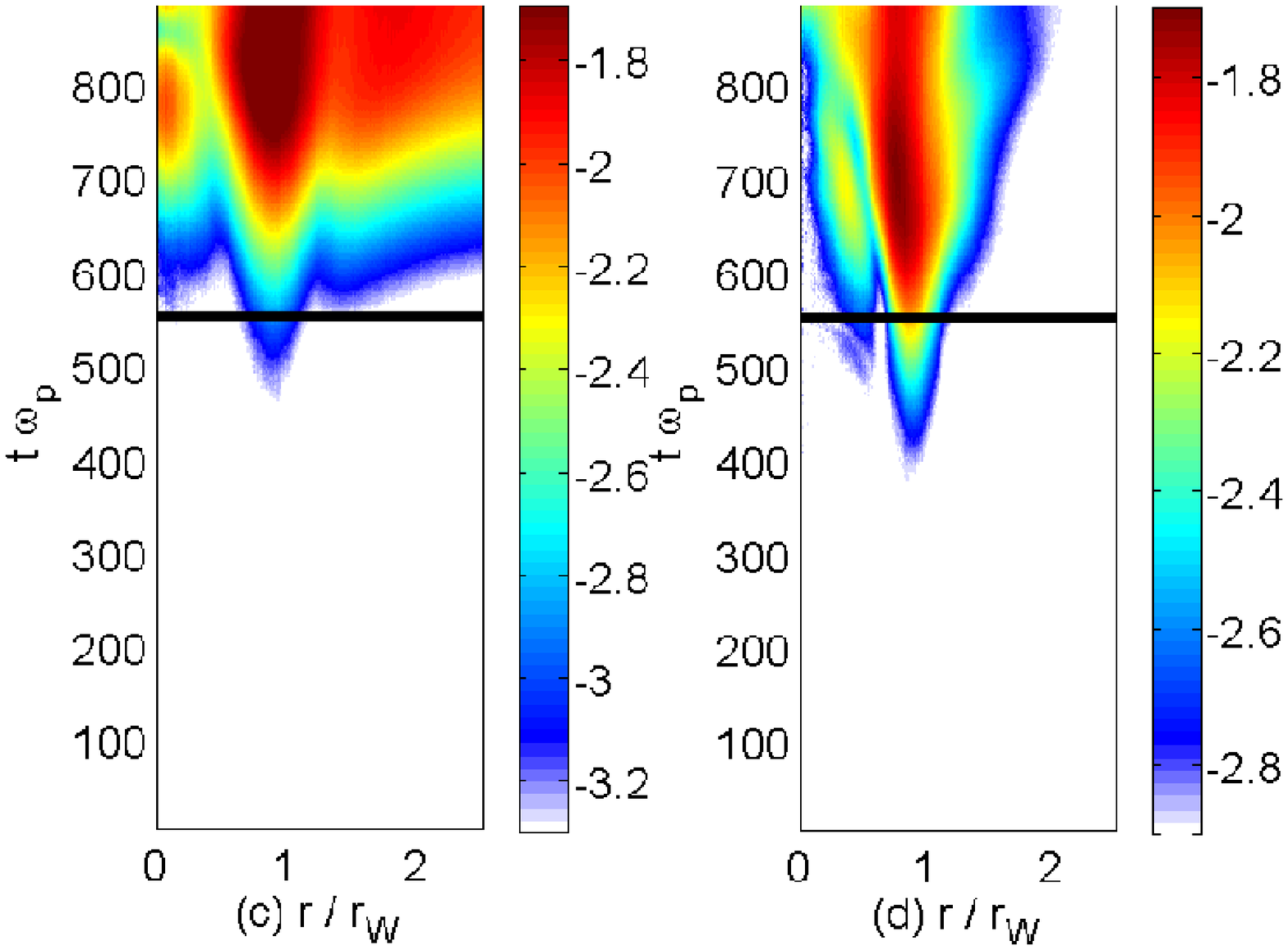}
\caption{(Color online) The azimuthally averaged electromagnetic fields ${\langle E_p^2 \rangle}^{0.5}$
(a), ${\langle E_z^2 \rangle}^{0.5}$ (b), ${\langle B_p^2 \rangle}^{0.5}$ (c) and ${\langle B_z^2 
\rangle}^{0.5}$ (d). The color scale is 10-logarithmic. The black horizontal line corresponds to
the time $T_2$.}\label{fig11}
\end{figure}

The saturation is demonstrated by Fig. \ref{fig11}, which compares the 
time-evolution of the mean amplitude of the azimuthally averaged fields ${\langle E_p^2 
\rangle}^{0.5}$, ${\langle E_z^2 \rangle}^{0.5}$, ${\langle B_p^2 \rangle}^{0.5}$ and 
${\langle B_z^2 \rangle}^{0.5}$. The supplementary movie 3 animates in time the spatial 
distributions of the in-plane electric field $E_p$ (a), the in-plane magnetic field $B_p$ (b) 
and the $B_z$. 
The in-plane electric field $E_p$ in Fig. \ref{fig11}(a) peaks initially at $r=r_W$. The build-up 
of this electrostatic field launches electrostatic waves into the cloud, which yields the amplitude 
peaks for $r < r_W$. The electric field that expands to larger $r > r_w$ and reaches $r \approx 2 r_W$ 
at $t=50$ is induced by the electrons that escape from the cloud, e.g. those with $r > r_W$ in 
Fig. \ref{fig3}(a). The electron current is not balanced by an ionic current and a radial electric field 
grows. The density of the escaped electrons decreases as they spread over the simulation box and the 
electric amplitude eventually drops below the threshold of Fig. \ref{fig11}(a). The position associated 
with the maximum of $E_p$ accelerates in time. This maximum coincides with the tip of the rarefaction 
wave. The electric field with the uniform amplitude 
of $10^{-2}$ behind the tip of the ion front roughly outlines the extent of the rarefaction wave. It expands to lower $r < r_W$ as the 
cloud is progressively eroded. The electric field component $E_z$ in Fig. \ref{fig11}(b) remains at 
noise levels, except for a peak at $t \approx T_2$. This peak coincides spatially and temporally with 
the growth of the rotational magnetowave, which is responsible for the $B_p \neq 0$ in Fig. 
\ref{fig11}(c). It is related to the displacement current $dE_z / dt$ connected to $B_p$. 

A comparison of the Figs. \ref{fig11} (c) and (d) confirms the observation from the supplementary
movie 3 that $B_z$ starts to grow to a large amplitude well before $B_p$. The latter grows fastest 
at around the time $t \approx T_2$, when $B_z$ saturates. The presence of a strong $B_z$ is apparently 
necessary for the growth of $B_p$, which supports our previous claim that $B_p$ is driven by a secondary instability. Figure \ref{fig11} reveals that $B_z$ is confined to within the rarefaction wave, 
while the magnetic field distribution $B_p$ reaches out into the dilute electron plasma outside of the rarefaction wave. It eventually reaches the boundaries and effects due to the periodic 
boundary conditions can no longer be neglected at the final time $T_F$. The development of the strong 
magnetic fields at $r\approx r_W$ coincide with a balancing of $K_r$ and $K_\rho$ in Fig. \ref{fig10}(b). 
This change in $A$ is partially due to the electron deflection by $\mathbf{B}$, which will contribute 
to an equilibration of $K_r$ and $K_\rho$.

We want to determine the source mechanism of the in-plane magnetic field $B_p$. The ambipolar electric 
field due to the rarefaction wave and the electric field induced by $B_z$ are both confined to the 
simulation plane. The electric fields and $B_z$ can excert a force on the electrons only in this plane 
and no current orthogonal to the simulation plane can be generated. Since we have no initial beams 
moving along the z-axis, we can rule out magnetic beam instabilities \cite{Review,Bret2,Bret}. One 
possibility to generate an electron flow along z is a coupling of the electron velocity components 
through the Lorentz factor. An electron acceleration in the simulation plane would change $v_z$. The 
fastest electrons in Fig. \ref{fig4}(a) have the mildly relativistic speed $v \approx 0.7c$ that gives 
a Lorentz factor $\Gamma \approx 1.4$. However, their number density is low and we could not expect 
them to drive strong in-plane magnetic fields. A possible mechanism for the growth of $B_p$ is the
electric field induced by $B_z$. The induced field is strong when $B_z$ saturates and it changes the
direction of the in-plane electric field vector by up to 5\% relative to the radius vector. The
plasma density is according to Fig. \ref{fig6}(b) circularly symmetric at $t=T_2$. A misalignment
of the in-plane electric field vector and the density gradient can give rise to elliptically
polarized magnetowaves \cite{Saleem,Saleem2}. 

\section{Discussion}

We have examined here the expansion of an initially unmagnetized circular plasma cloud into a 
vacuum by means of a 2D3V PIC simulation. The purpose of our study has been to test, if a thermal 
anisotropy-driven Weibel instability (TAWI) can also develop in curved rarefaction waves. Previous
studies considered planar plasma slabs that expanded under their own thermal pressure \cite{Thaury}
and observed the growth of strong magnetic fields in the rarefaction wave, unless a shock suppressed 
the TAWI \cite{SarriPRL,SarriNJP}. 

A thermally anisotropic electron distribution develops due to two reasons. Firstly the slow-down of 
the electrons in the ambipolar electric field implies that their thermal pressure is decreased along 
the density gradient. This mechanism occurs in planar and curved rarefaction waves. Secondly, the 
inelastic reflection of an electron by the expanding rarefaction wave implies that it looses momentum 
along the electric field direction. This momentum is transferred to the accelerating ions. The initial 
plasma distribution used in the Refs. \cite{Thaury,SarriPRL,SarriNJP} is a plasma slab of infinite 
extent in one direction, which has a finite width in the second direction. This plasma slab is immersed 
in a vacuum or in a low-density plasma. Both plasma boundaries are parallel. This implies that every 
reflection of the electron will decrease the same momentum component, namely the component that is 
aligned with the antiparallel density gradients of both rarefaction waves. The anisotropy increases 
with every electron reflection. This mechanism can not occur in a circular rarefaction wave, since 
here consecutive electron reflections tend to decrease different momentum components. Our study can 
thus test if consecutive electron reflections are necessary to achieve a TAWI. 

If the slowdown of the electrons in the ambipolar electric field of a plasma with a spatially 
non-uniform density is enough to trigger the TAWI, then this instability may play a role in the
magnetization of astrophysical plasma \cite{Schlickeiser2} besides the filamentation instability \cite{Medvedev} and
cosmic ray-driven instabilities \cite{Bell}. The magnetic energy that can be generated 
by the TAWI is a fraction of the electron thermal energy, here about 1\%. This energy density is well below that
of the filamentation instability, which can reach about 10-20\% of the kinetic energy density of 
leptonic beams if they are cold. However, the TAWI requires only hot electrons and a density gradient 
to grow, while the filamentation instability is triggered by relativistic particle beams that we may
find only close to plasma shocks. The TAWI could thus magnetize much larger intervals in
energetic astrophysical jets, such as GRB jets, which are thought to have a nonuniform plasma
density and relativistically hot electrons. The TAWI may also generate a seed magnetic field
in the turbulent plasma close to SNR shocks and speed up the development of the much more powerful 
cosmic ray-driven instabilities. 

Our findings are as follows. The electron expansion results in a strong radial electrostatic field. 
It counteracts a further electron expansion and accelerates the protons into the radial direction. 
A rarefaction wave emerges, which is characterised by an increase of the proton mean speed and by a 
decrease of the proton density with an increasing radius. The expansion of a rarefaction wave is 
self-similar \cite{Schamel,Grismayer,Lyutikov}. The absence of an ambient medium has prevented the 
formation of a shock and the protons could reach a speed exceeding 250 times their thermal speed.

The fields within the rarefaction wave have initially been electrostatic. Electrons lose momentum 
to the accelerating protons only in the radial direction, which implies that their radial thermal 
pressure decreases more quickly than the azimuthal one. A thermal anisotropy-driven Weibel 
instability (TAWI) \cite{Weibel} grows once the radial extent of the rarefaction wave is sufficient 
to accommodate a wavelength of the unstable modes \cite{Thaury} and once the thermal anisotropy 
between the radial and the azimuthal direction became sufficiently large. The TAWI results in the 
aperiodic growth of a transverse magnetic (TM) wave. This TM wave induces an electric field that 
is added to that of the electrostatic ambipolar electric field of the rarefaction wave. The total 
electric field is thus no longer parallel to the plasma density gradient.

A misalignment between the electric field and the plasma density gradient can drive circularly or elliptically polarized magnetowaves \cite{Saleem,Saleem2}. We propose that 
these waves are responsible for the in-plane magnetic field we observe in our simulation, which 
can not be explained in terms of the TAWI. The magnetic fields of this wave reach an amplitude 
that is comparable to the waves due to the TAWI.

Our findings can be verified in an experiment, where the absorption of a laser pulse heats up the electrons in a thin
wire \cite{Quinn}. Some electrons will leave 
the now positively charged wire and build up an ambipolar electric field. This ambipolar field 
accelerates the ions through the transverse normal sheath acceleration (TNSA) mechanism \cite{Snavely,Gremillet1} and it results
in the formation of a rarefaction wave. The ablation of a wire generates a cylindrical rarefaction
wave. Our circular plasma cloud corresponds to an idealized cross section of the wire, as long as it is
sufficiently far away from the laser impact point \cite{Quinn}. The cross section
is idealized, because we do not consider ionization processes that require more elaborate numerical
codes \cite{Gibbon,Gremillet2}. The plasma dynamics outside of the wire should be comparable in the 
simulation and experiment. The strong magnetic fields that develop in the rarefaction wave should be 
detectable. It is thus possible to investigate 
systematically the TAWI in a controlled laboratory experiment and to determine the strength and 
life-time of the magnetic field. The interplay of the TAWI with the secondary magnetowave we have observed here and
other instabilities, such as the Rayleigh-Taylor instability \cite{Piriz1,Piriz2}, can also be examined. The experiment can reveal the topology of the magnetic fields in 
a three-dimensional setting, which is currently computationally too expensive. 

Acknowledgments: This work was supported by Vetenskapsr\aa det (DNR 2010-4063), EPSRC (EP/E035728/1, 
EP/C003586/1, EP/D043808/1), Consejeria de Educacion y Ciencia (ENE2009-09276), the Junta de Comunidades de 
Castilla-La Mancha (PAI08-0182-3162), SFI (08/RFP/PHY1694). Computer time and support was provided 
by the HPC2N in Ume\aa \, and by the ICHEC in Dublin.

\end{document}